# An Autonomous Self-Incremental Learning Approach for Detection of Cyber Attacks on Unmanned Aerial Vehicles (UAVs)

Yasir Ali Farrukh, and Irfan Khan, Senior Member *IEEE*

*Abstract*— **As the technological advancement and capabilities of automated systems have increased drastically, the usage of unmanned aerial vehicles for performing human-dependent tasks without human indulgence has also spiked. Since unmanned aerial vehicles are heavily dependent on Information and Communication Technology, they are highly prone to cyber-attacks. With time more advanced and new attacks are being developed and employed. However, the current Intrusion detection system lacks detection and classification of new and unknown attacks. Therefore, for having an autonomous and reliable operation of unmanned aerial vehicles, more robust and automated cyber detection and protection schemes are needed. To address this, we have proposed an autonomous self-incremental learning architecture, capable of detecting known and unknown cyber-attacks on its own without any human interference. In our approach, we have combined signature-based detection along with anomaly detection in such a way that the signature-based detector autonomously updates its attack classes with the help of an anomaly detector. To achieve this, we have implemented an incremental learning approach, updating our model to incorporate new classes without forgetting the old ones. To validate the applicability and effectiveness of our proposed architecture, we have implemented it in a trial scenario and then compared it with the traditional offline learning approach. Moreover, our anomaly-based detector has achieved a 100% detection rate for attacks.**

## I. INTRODUCTION

With the rapid advancement in the field of automation, Information & Communication Technology (ICT), every sector in the world is now shifting towards autonomous and digitalized systems for ease and efficient operation. Similarly, aerospace, a branch of science related to aviation and space flight, has also adapted this recent development and has embedded artificial intelligence (AI) in its operation. This has resulted in significant advancements in Unmanned Aerial Vehicles (UAVs), a pilotless onboard craft also known as a drone. UAVs can autonomously fly and be functioned remotely[1], and due to their high mobility, it has proven to be a great asset in many applications. In the early period, UAVs were utilized for military purposes only. However, they are recently being used in every domestic sector, including filmmaking, agricultural plant protection, goods delivery, or even health care.

Deployment and usage of UAVs have increased drastically over time[2]. According to the Federal Aviation Administration (FAA), it has been estimated that a total of seven million UAVs are being operated in the United States alone [3]. Moreover, it has been anticipated that by the end of 2036, the total revenue of

UAVs will be more than $30 billion. Recently, UAVs have been in the spotlight due to their effective operation independent of a human pilot. This ease and convenience are proving to be extremely useful. However, at the same time, it brings new challenges and hurdles, e.g., the operation of UAVs can be unintentionally or intentionally interrupted, causing harm to people or their privacy.

Moreover, the major concern for UAV is cyber security. With time, new cyber attacks are being developed, and the frequency of such attacks is increasing day by day[4]. With such attacks, the integrity of UAVs can easily be exploited, causing enormous damage.

Cyber security is one of the hot topics for UAVs, and much research is undergoing for the development of methods and techniques for securing UAVs. As a large number of sensors are used in UAVs for collecting and processing data before their transmission to other networks, this attracts the attackers to exploit UAVs' networks, resulting in abnormal operation[5]. To mitigate these threats, many intrusion detection systems (IDS) have been proposed, and a majority of the existing solution is based on Machine Learning (ML). IDS based on ML monitors network traffic and analyzes to detect malicious activity and suspicious attacks. IDS aims to identify and detect different cyber attacks by examining collected data to take necessary actions to avoid any mishap. Primarily, there are two forms of IDS according to the detection strategies of network attacks[6]. The first one is signature-based detection which compares collected data with a knowledge base of known attacks or intrusions. However, this method cannot detect new attacks, which were not present at the time of model training[7]. The second one is anomaly-based detection that compares collected data with a trained model of standard user behavior and identifies data instances that deviate from the normal behavior as an anomaly[6]. As a result, this approach can identify unseen attacks but cannot classify the type of attack[7]. However, in this paper, we have combined both techniques to develop an architecture capable of classifying both seen and unseen attacks. Several public intrusion detection datasets have been published over the years; however, there is still a need for a more robust and accurate dataset relating to a specific application. The literature review shows that much work has been proposed using the KDD Cup dataset assembled in 2009, resulting in an outdated dataset as network traffic present in it no longer corresponds to today's usage. A more recent and realistic dataset, CSE-CIC-IDS-2018 is produced in 2018 and must be utilized until application-specific, and more robust datasets are introduced. For our research work, we have utilized this dataset to cross-validate our proposed architecture, and therefore, our literature review only comprises work including the CSE-CIC-2018 dataset.

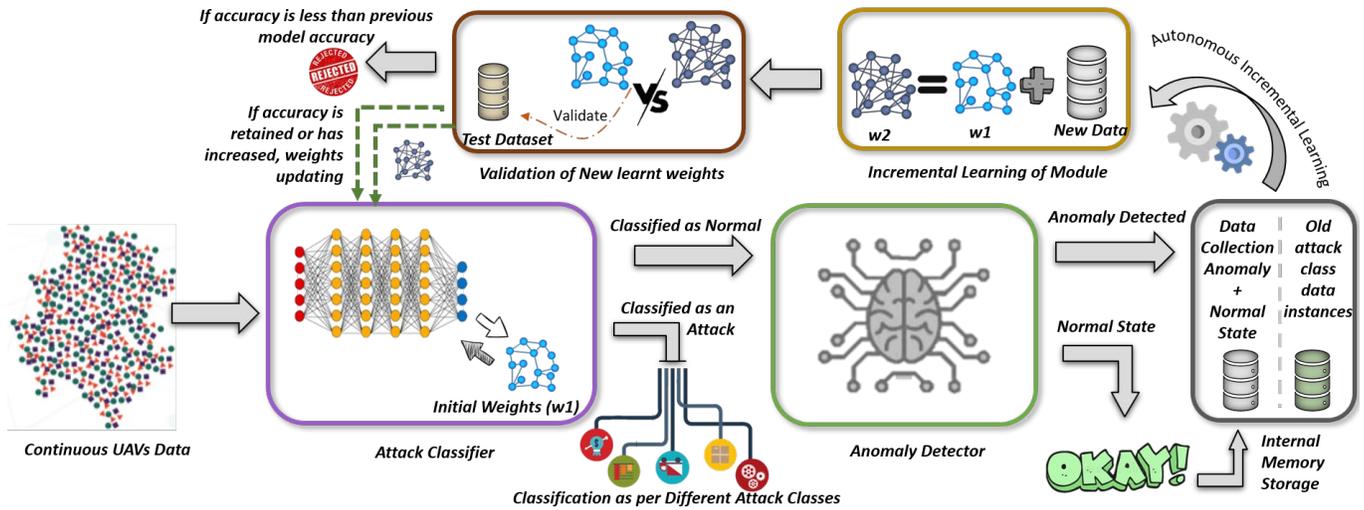

Fig. 1. Proposed architecture for autonomous self-incremental learning approach. The architecture comprises of four components; Attack Classifier, Anomaly Detector, Internal memory storage and Incremental learning module. Attack classifier is based on supervised machine learning approach whereas anomaly detector is based on semi-supervised machine learning approach. The proposed architecture is independent of any human interference.

A signature-based IDS using an aggregator model integrating four network architecture models has been proposed in [9]. The model comprises of restricted Boltzmann machine, a deep feed-forward neural network, Recurrent Neural Networks (RNNs), Long Short-term Memory (LSTM), and a Gated Recurrent Unit (GRU). The highest accuracy (100%) was obtained with this aggregator model for the DoS, DDoS, and brute force attack types. However, the author does not present the result outcomes of the combination of attacks. Similarly, Ferrag et al in [10] evaluated seven deep learning algorithms, including RNNs, deep neural networks, restricted Boltzmann machines, deep belief networks, Convolutional Neural Networks (CNNs), deep Boltzmann machines, and deep autoencoders on the CSE-CIC-IDS-2018 dataset but lacks experimental details.

Moreover, conventional classifiers like Random Forest (RF), Decision Tree (DT), k-NN, Naive Bayes, and Single Vector Machine (SVM) were evaluated in [11]. The overall result suggests that RF and DT had the best performance among all. However, insufficient information has been provided in the paper regarding data processing. Another approach using Attention Mechanism (AM) with LSTM is proposed in [12]. The author evaluates the accuracy and recall rate of 6 categories compared to traditional and deep learning approaches. However, this paper also lacks proper details regarding data cleaning.

In terms of anomaly-based detection, a two-level anomaly detector using Auto Encoder (AE) is proposed in [13]. The author achieved a detection rate of 99.2%. However, the false alarm rate of the DoS attack is observed to be high as compared with similar methods. Similarly, Mezina et al. in [14] have utilized U-Net-based and temporal convolutional network CNN. The overall accuracy achieved by the author on the CSE-CIC-IDS-2018 dataset is 97% and 94%, respectively.

Although much work has been done in the domain of IDS claiming an accuracy near to 100%, there is still a gap for identifying and classifying new unseen attacks as there is no one solution for classifying known and unknown attacks. However, in [a], the author propsed a two-stage framework for detecting unknown and old attacks but lacks classification of these new attacks. Therefore, in this paper, we have proposed a novel architecture that is capable of performing this task autonomously. As per the authors' best knowledge, no such architecture has been proposed in the existing literature and is being presented for the first time in this paper. We propose a model combining signature-based IDS and anomaly-based IDS to detect known and unknown attacks. We first pass the new data through signature-based IDS, which classifies the attacks as per the known classes in the proposed architecture. Suppose the signature-based IDS classifies an instance of data as a normal class. In that case, it will then be passed onto the anomaly detector, ensuring that this is a normal class and no new attack has been initiated. If the anomaly detector detects the data instance as an anomaly, it is alerted, and new attack data instances are being collected in the internal memory. This new attack data is then mixed with a chunk of old training instance data and is fed to the IDS for incremental learning of this new class. After completing this incremental learning, our signature-based classifier is now capable of classifying the new attack class. We are continuously updating our classifier through this approach without any intervention from humans. The proposed architecture is shown in Fig. 1. The main contributions of this article are summarized below.

1) An autonomous class incremental learning architecture has been proposed for the first time to address the problem of sensing and tracking new/unknown attacks.

2) A semi-supervised based anomaly detector is developed, having an attack detection rate of 100%. It is also then compared with other recent state-of-the-art approaches.

3) A brief discussion is presented regarding the different approaches for class incremental learning and their challenges. Later, incremental learning is implemented and evaluated.

The rest of the paper is arranged as follows; Section II provides a brief description regarding preliminaries. In Section III, a detailed overview of the proposed architecture and adopted methodology has been discussed. Experimental results and outcomes are given in Section IV, and Section V concludes this research paper.

## II. PRELIMINARIES

### A. Signature Based Intrusion Detection

A signature-based intrusion detection system uses supervised classification algorithms for the identification of the different categories of attacks. In the classification algorithm, the model learns from the available labeled dataset and then predicts new data following its training. In simple words, a discrete output function $y$ is mapped as per the input features or variables $x$.

The main objective of the classification algorithm is to identify the given data points as per the given categories and are mainly used for the prediction of categorical data. Classification algorithms can be divided into two types of classification: binary and multi-class classifications. In binary classification, there are only two categories, whereas, in multi-class classification, there are more than two categories to identify. In terms of working principles, it can be broadly divided into two main categories: a linear model and a non-linear model. A linear classification model identifies classes or categories based on the value of a linear combination of the features that is the input value, while non-linear models apply non-linear methods such as quadratic discriminant analysis (QDA), regularized discriminant analysis (RDA), and mixture discriminant analysis (MDA)[15]. In our work, we have utilized a non-linear approach. Examples of linear and non-linear models are logistic regression, support vector machine, random forest classifier, naïve Bayes & neural network, respectively.

### B. Anomaly Based Intrusion Detection

Anomaly detection is the method for determining an unusual pattern or point in the given dataset[16]. It can be broadly categorized into outlier detection and novelty detection. The key difference between outlier detector and novelty detector is that the training dataset of the outlier detector contains outlier, which is different from normal instances, whereas in novelty detector, the training dataset only contains normal condition data and it works on the principle of detecting any data that deviates from the normal condition. Both of these approaches are widely used in the field of intrusion detection. Based on working principles, these detection models can be broadly categorized into three categories: Unsupervised, supervised, and semi-supervised. Unsupervised anomaly detectors identify anomalies in an unlabelled dataset under the assumption that the majority of data instances are normal and instances that are least similar to the majority of data are identified as an anomaly. In a supervised approach, labeled data is fed to the model for training, and then it predicts the new data as per the training. The difference between this approach and the binary classification approach is the inherent nature of unbalanced outlier detection. A semi-supervised anomaly detector develops a model representing normal condition from a dataset containing only a normal state and then predict the data as per the learning. In our approach, we have utilized a semi-supervised approach, which belongs to a novelty detection category.

### C. Incremental learning

Incremental learning is the ability of the model to combine various learning objectives or tasks seen over time into a coherent whole[17]. As data is being continuously generated over time, new tasks or classes emerge regularly. Therefore, the need for updating the model to facilitate these additions is crucial. One approach can be to train a new model from scratch, but this approach is infeasible due to several issues like computational and storage problems. Therefore, a need for continuous/lifelong learning arises. Incremental learning is also a part of continual learning, as the main objective of continual learning is to refine and accumulate knowledge over long timespans. Continual learning can be divided into three subcategories:

    a) Online Learning
    b) Transfer learning
    c) Incremental learning

The online learning approach aims to update and learn new data instances of known classes[18]. It updates the model as per the new data, unlike traditional machine learning (also known as offline learning), which runs in batch learning, where training data is provided in advance to train the model. Offline learning requires the entire training dataset to be available before training, leading to several drawbacks like low efficiency and poor scalability[18]. Through online learning, we can overcome these drawbacks as it can be updated instantly for new data instances. In short, online learning can update the model as per the new data, but it cannot learn new classes or tasks. It can only update the pre-trained model based on new data instances.

Transfer learning is a machine learning approach in which a model trained for some tasks is reused as starting point for another task's model[19]. Basically, transfer learning is an optimization that helps in performance improvement and rapid progress while training the model on the second task. It is related to problems like concept drift and multi-task learning. Transfer learning only works if the model features learned from the precedent model are general. It is widely popular in deep learning, where pre-trained models are utilized as a starting point for natural language processing and computer vision tasks, as this type of application requires huge computational and time resources. In short, transfer learning allows transferring the knowledge learned from one task to another different task[20].

Incremental learning is an approach similar to online learning as it aims to update the model to learn new classes or tasks from new data while retaining the old ones too. In most incremental learning, task or classes are given in sequential order so that after each training, a model can be cross-validated on the test data set of all the classes. This determines if the model has retained the previous knowledge of all the classes and the updated knowledge of newly learned classes or tasks [21]. The main challenge for incremental learning is to learn new classes in a way that prevents forgetting of the previously learned classes[22]. As we do not have access to previous original data, therefore updating the model on new data instances will mold the model to perform well on new tasks/classes and will have a significant drop in performance for old classes or tasks. In short, class incremental learning updates the model to learn new classes or tasks while retaining the knowledge of the previous classes or tasks, simply it will upgrade the model to work well on both new and old data classes or tasks.

The problem of performing well on new tasks or classes while performing poorly on the old ones is termed as catastrophic forgetting[23]. A lot of research is being held in this domain, and there is a tradeoff between rigidity that is performed on old tasks and plasticity, which is performed on new tasks. For overcoming catastrophic forgetting, a model must show the ability to acquire new knowledge and refine the old learning based on the new data instances. Moreover, a model should be capable of preventing interference with the existing learning due to the data of new classes or tasks. The optimal point for a system to be plastic to integrate new information and retain stability is known as the stability-plasticity dilemma.[24]. To overcome this issue, three main strategies are being widely used.

a) Methods based on external memory
b) Constraint-based methods
c) Methods based on model plasticity

Methods based on external memory use some portion of old data used for the initial training of the model while updating the model with the new classes or task data. Rebuffi. et.al in [25] has presented a method based on a similar approach in which the author specifies a limited amount of space for storing previous data. In this way, the model is incrementally learning new classes while combining some instances of old class data. However, the performance of this approach is still lower than what the model can achieve if it is trained completely again by the offline learning technique. Another approach is to store class distribution statistics instead of retaining a portion of previous data. This approach was presented in [26], [27]. With this approach, a generative model can produce on-the-fly old classes data. However, this technique is very reliant on the quality of the generative model.

The constraint-based method adapts the approach for forcing the updated model $M^t$ to be similar to its previous version that is $M^{t-1}$. Several methods are utilized to achieve this goal. However, all these methods have to maintain a balance between rigidity (encouraging similarity between $M^t$ and $M^{t-1}$) and plasticity (allowing enough slack $A^t$ for to learn new classes). These methods can be separated into three categories; based on activations, based on weights, and based on gradients. In [28], the author has proposed a method using knowledge distillation from[29], that is, molding the model to mimic the activations of its previous version. This approach reduces the forgetting of old classes, such that if the same data is given to both the models the base probabilities of them should be similar and distillation loss will simply be binary cross-entropy. The second approach of using weights is to minimize the distance between the old and new weights of the model. Kirkpatrick et al in [23] further suggested modulating the regularization according to the importance of neurons. Neurons that are important for old classes must not be changed in the new model. This approach of refinement of neurons weights is also presented in [30], [31]. The last approach of using gradients comprises the idea that the loss of a new model should be lower or equal to the loss of an old model on samples stored in memory, that is, old instances of test data. Simply this approach forces the new model to go in the same direction as the previous version of the model would have. This approach is introduced in [32].

The last strategy based on model plasticity is to modify the network structure to reduce catastrophic forgetting. In this strategy, a new neuron can be added to the current model and dedicated to learning the new class. Some old neurons in the network which have high importance can also be frozen to avoid forgetting. Moreover, several methods also exist to uncover the subnetwork and sparsify the whole network with regularization. Golkar et al in [33] have adopted such an approach. It is worth noting that techniques based on such an approach assume to know on which task they are being evaluated.

In short, there are many approaches for performing incremental learning, and many authors are using a combination of approaches to achieve their goals. A brief overview of the categorization of continual learning is shown in fig. 2. In our work, we have employed incremental learning as well as offline learning approach to have a better understanding between achieved results.

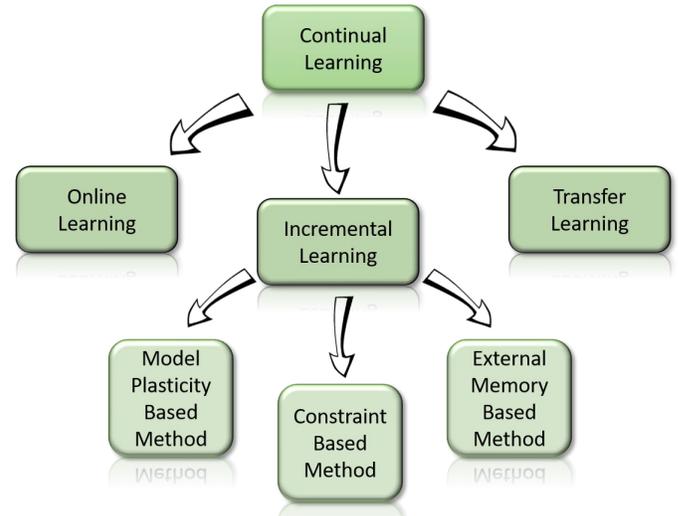

Fig. 2. Categorization of Continual learning. Incremental learning is categorized further based on its operational strategies.

### D. CSE-CIC-IDS2018 Dataset

CSE-CIC-IDS2018 data is the outcome of the collaborative project of the Communications Security Establishment (CSE) and The Canadian Institute for Cybersecurity (CIC) that utilizes the profiles notion to generate dataset systematically. However, it is not a new project but a part of an existing project that produces data in a scalable manner. Similar to the CIC-IDS2017 dataset, the CICFlowMeter tool is being used to extract 80 network features. This dataset includes a variety of intrusion attacks along with their abstract description for application and protocols. In addition, this dataset includes seven unique attack scenarios and their network traffic and logs files. The dataset is distributed among ten days of network traffic, resulting in 16,233,002 instances of data. However, only 17% of the dataset is attack instances. The percentage distribution of the CSE-CIC-IDS2018 dataset can be seen in Table I. The dataset is divided into 10 CSV files that can be downloaded from [34]. This dataset is the most recent available dataset in terms of network intrusion attacks.

TABLE I
DISTRIBUTION OF CSE CIC-IDS2018 DATASET

| Traffic Type | Distribution |
|---|---|
| Benign | 83.07 % |
| DDoS | 7.78 % |
| DoS | 4.03 % |
| Brute Force | 2.35 % |
| Botnet | 1.763% |
| Infiltration | 0.996% |
| SQL | 0.006% |

## III. PROPOSED ARCHITECTURE AND METHODOLOGY

Several models and approaches have been proposed in the literature about intrusion detection of known attacks, but the problem of detecting new and unknown attacks persist. However, there are approaches like[13] that have used anomaly detectors to distinguish between normal and attack instances. Although these types of models using anomaly detectors can distinguish new attacks, they lack identification of attack type, which can be quite useful in taking immediate steps to avoid any damage. Keeping this need in mind, an autonomous self-incremental learning approach has been proposed capable of classifying known attacks along with the classification of unknown and new attacks as well. The proposed architecture is shown in fig 1. Through this approach, a UAV can automatically detect and update its IDS without the need for human interference. The proposed architecture comprises two main components which work in sequential order. The first component is the attack classifier which uses a supervised machine learning approach for identifying the different classes of attacks. On the other hand, the second component is the anomaly detector which uses a semi-supervised approach to distinguish between normal data and abnormal data. Other than that, there is internal memory storage that stores some proportion of known attack data instances utilized while training and testing dataset instances required for the incremental learning of the attack classifier. In addition to that, there is also an incremental module that trains the model and validates whether these new instances are attacks or not. The overall flow of the proposed architecture is summarized in Algorithm 1.

---

**Algorithm 1** Autonomous Self-Incremental Learning

---

Network Features ◄——— Stores UAVs Network Data after Preprocessing
Send these Network Features to Trained Attack Classifier
Prediction ◄———Attack Classifier classifies the data as Normal Traffic or Specific Attack Class
**if** Prediction is Attack:
    Initiate Alarm for preventing system from Identified Threat
**else:**
    Send Features to Anomaly Detector (LOF):
    Pred_Anomaly ◄———LOF classifies the data as normal or anomaly
    **if** Pred_Anomaly is Normal:
        **if** Normal Instances in Internal Memory is < N number of data instances
            Stores in Internal Memory Storage as Normal
        **else**
            Overwrite the old Normal data in Internal Memory Storage
        **end**
    **else:**
        Stores in Internal Memory Storage as New Attack and initiate warning
        X_train_new_batch ◄——— Combine Normal, New Attack and Old Attack
                            Data Instances in Proportion
        Sends X_train_new_batch to incremental learning module
        Perform Class incremental learning on Attack Classifier
        Performance_new ◄——— Calculate Evaluation Metrics on stored test dataset
        **if** Performance_new < Performance Old - X%:
            Discard new model obtained via Incremental Learning
        **else:**
            Attack Classifier model ◄——— Replace Old model with New
                              model weights
            Updated Model can classify newly identified threat
        **end**
    **end**

**end**

---

### A. Attack Classifier

Neural Network has been incorporated as an attack classifier to identify different classes of attacks. However, conventional methods like Random Forest (RF) and Decision Tree (DT) classifier also perform quite well but lack the ability of incremental learning. Since one part of our architecture is dependent on incremental learning, therefore, a neural network was given preference. However, RF has also been evaluated, and an alternate method for offline learning is also assessed side by side. Among several neural network approaches, Bidirectional Long short-term memory (BiDLSTM), an extended version of LSTM is utilized. Conventional LSTM is an extension of Recurrent Neural Network (RNN), developed to avoid the long-term dependency issue, unlike RNN, it can retain data for long periods. BiDLSTM enhances standard LSTM to improve the model performance[35]. It trains two LSTM on the input, the first one trains on the input data while the second one trains on the reversed replica of the input data. Doing so, more meaning is added to the model, and results are achieved faster. The main reason for choosing BiDLSTM was that the data could be processed in two directions using a forward hidden layer and a backward hidden layer. Since the CSE-CIC-IDS2018 dataset is a sequential instance that is time-dependent, the LSTM approach tends to perform well on such problems[36]. The selected parameters for BiDLSTM are shown in Table II.

### B. Anomaly Detector

A Semi-supervised approach has been adopted for anomaly detection. A novelty detector was preferred instead of outlier detectors, as our main goal is to identify any data that differs from a normal instance. For this purpose, the Local Outlier Factor (LOF) was utilized as LOF computes the local density deviation of a given point to its neighbors which is quite effective in determining the outliers.

| Parameter | Description |
|---|---|
| Number of Layers | 5 |
| Layer 1 | BiDLSTM (None,1,96) |
| Layer 2 | Dropout-0.2 |
| Layer 3 | LSTM (None, 32) |
| Layer 4 | Dropout-0.2 |
| Layer 5 | Fully Connected (None,11) |
| Activation Function | Softmax |
| Loss Function | Categorical cross entropy |
| Optimizer | Adam & lr=0.001 |
| Batch Size | 1024 |
| Epochs | 50 |

### C. Internal Memory Storage

In our proposed architecture, we are utilizing a mixture of incremental learning approaches that combine external memory and constraint-based methods. To achieve an optimal result for our model, we are storing a small proportion of 10000 training points for each attack class so that these old attack classes can be combined with the fixed proportion of the new attack class and normal class for incremental training of the model. Moreover, the test dataset is also stored in internal memory which will aid in evaluating the new attack class. This is the same test dataset that was utilized in evaluating the performance of the models.

### D. Incremental Learning Module

The function of the incremental learning module is to gather the data from internal memory and newly receive data in proper proportion and then feed it to the model for incremental learning. In our approach of incremental learning, we have utilized a combination of memory and constraint-based method. We are reinitializing the weights of the model as per the previous model and adding neurons in the last layer as per the increment of the new attacks. After the initialization of weights, training is performed on a new batch of training data. To validate whether the detected attack is an actual attack or not, this module cross-checks the recall of the previous model and the currently trained model. If the results are in the range of -2% or better, then the model is updated, otherwise, it is disregarded as it is degrading the performance of the model and this attack class was just a false alarm.

### E. Cleaning and Preprocessing of Data

Cleaning and preprocessing data is the initial step for solving any given problem. Since the CSE-CIC-IDS2018 dataset is comprised of 10 CSV files, they were combined into a single file and read. The CSV file of Thursday 20-02-2018 has four additional features which were removed as they were not present in other CSV files. The first step for cleaning and preprocessing was to remove any infinity and nan values present in the dataset. Afterward, duplicate data was also removed as it would not help in learning the model. Another problem observed during preprocessing of data was that there were some string values available in the dataset that were also removed. Upon exploring, it was seen that the header column was also the part of data for several entries. After processing data, all attacks instances were separated, and all normal instances were separated. The number of attack instances after preprocessing is shown in table III.

| Attack Type | Number of Data Instances |
|---|---|
| DDoS attacks-LOIC-HTTP | 575364 |
| DDoS attack-HOIC | 198861 |
| DoS attacks-Hulk | 145558 |
| Bot | 144535 |
| Infilteration | 139805 |
| SSH-Bruteforce | 94048 |
| DDoS attack-LOIC-UDP | 1730 |
| Brute Force -Web | 555 |
| Brute Force -XSS | 228 |
| SQL Injection | 84 |
| DoS attacks-SlowHTTPTest | 57 |
| FTP-BruteForce | 53 |

Since the maximum number of attack instances are for DDoS attacks is 575364; therefore, the same instance of Benign data points was also extracted. This fixed proportion of data instances is kept to avoid class imbalance problems. After combining both attack and normal data instances, they were split into train, test, and validation set with a ratio of 50:30:20, respectively. Train and validation sets were oversampled using Borderline Synthetic Minority Oversampling Technique (SMOTE) to avoid imbalance classification problems. Suppose the data set is not handled properly. In that case, the resulting outcome will be a biased trained model giving an edge to the class having the majority of data instances and might overlook the minority class[37]. After Balancing each data instance through Borderline SMOTE, two random attacks were omitted from the dataset to be used in two scenarios. The attack classifier was trained on 12 classes, and the anomaly detector was trained on only normal class data instances. An overview of data cleaning and pre-processing is also illustrated in Fig. 3.

### F. Dimensionality Reduction

The number of features available in a dataset is known as the dimensionality of the dataset. It is directly proportional to the computational power and time required for training the model[38]. This issue is often termed as a curse of dimensionality. To overcome such issues, statistics and different reduction techniques are used for data visualization. The method utilized in this research for feature reduction in Mean Decrease in Impurity (MDI). The MDI is a method that measures the importance of features in the evaluation of a target variable. It calculates an average total decrement in node impurity, weighted by the ratio for each feature reaching that particular node in a different decision tree.

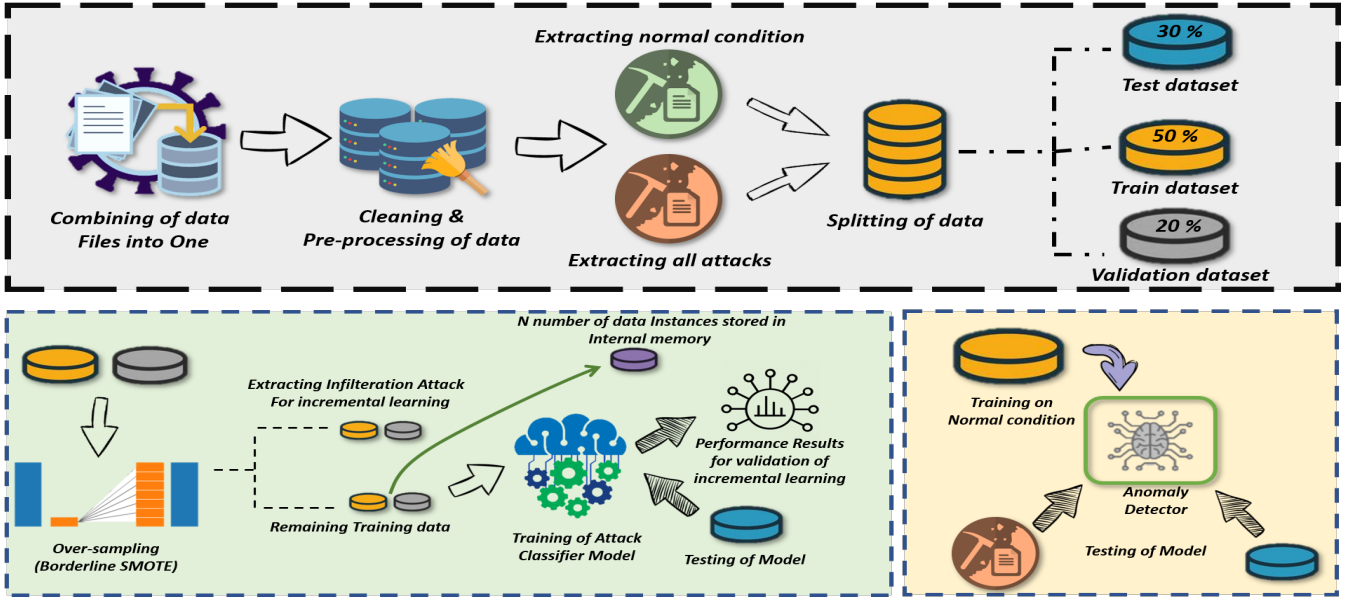

Fig. 3. An overview of data cleaning and preprocessing for the proposed architecture. All CSV files of CSE-CIC-IDS2018 data were compiled together and cleaned. All data instances of the dataset were not utilized, moreover data was balanced using borderline smote after splitting into train, test and validation dataset. The box at the top represents general preprocessing, after which data was processed as per attack classifier and anomaly detector. The right box is for anomaly detector whereas one is for attack classifier model.

Therefore, higher MDI is an indication that a particular feature has high importance for a particular task. The MDI index(G) is defined in (1)

$$G = \sum_{i=1}^{n_c} p_i(1 - p_i) = 1 - \sum_{i=1}^{n_c} p_i^2 \qquad (1)$$

where $n_c$ is the number of classes in the target variable and $p_i$ is the ratio of this class.

The decrease in impurity $I$ is then defined in (2)

$$I = G_{parent} - P_{split1}G_{split1} - P_{split2}G_{split2} \qquad (2)$$

where $P$ is the data proportion of each split that takes up the relative parent node.

This approach is utilized in the proposed architecture as it gives the mean decrease impurity of each feature based on multi-class classification. After analyzing the result and after some trial runs it was found out that the most optimal number of features for the classification of attack classes is 20. MDI score of the top 15 features of the dataset is shown in fig 4.

### G. Standardization of Dataset

Data standardization is very crucial for the training of a model as it can significantly impact the outcome of the training model. There are many approaches available for data standardization. The approaches utilized in our architecture are Standard Scaler and Min-Max scaling. Standard Scaler transform the data set, keeping mean 0 and variance 1. Standard Scaler is utilized for attack classifier. It is calculated as

$$Standard\ Scaler = \frac{x_i - \bar{x}}{\sigma} \qquad (3)$$

Min-Max Scaling transforms the dataset between a scale of -1 and 1. This method is utilized in an anomaly detector since the data set of the anomaly detector only contains normal condition data for training. It is calculated as

$$x' = \frac{x - \min(x)}{\max(x) - \min(x)} \qquad (4)$$

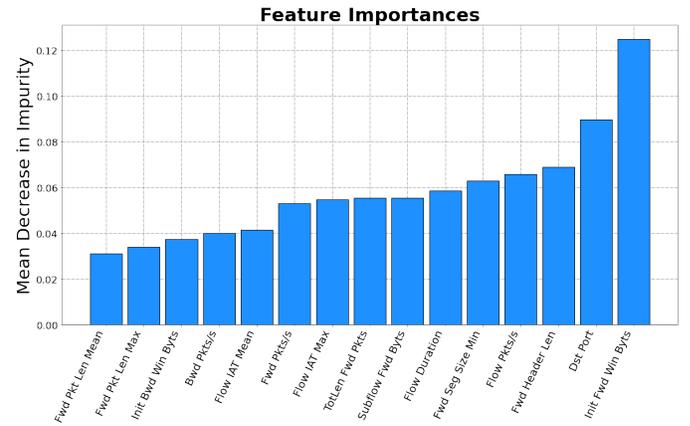

Fig. 4. Features having significant importance in distinguishing attack classes. Among 80 features, this figure represents the top 15 features in terms of mean decrease in impurity.

## IV. EXPERIMENTAL RESULT AND OUTCOME

The proposed autonomous self-incremental learning approach was trained and evaluated using Google Collaboratory. We have omitted one random attack from our train and validation dataset for performing incremental learning scenarios, assuming that this attack is new and unknown yet. After the removal of the attack, training was performed for our attack classifier on 12 different classes. However, we do not need any such thing for anomaly detectors as it only trains on a dataset containing a normal state.

We separated attack and normal instances from the training and validation dataset and fed all normal data instances to the LOF model for training the anomaly detector. For testing purposes, we tested our model on two dataset chunks. The first one is attack instances that were separated from the training and validation dataset, on which our model performs quite well, detecting attack instances with a 100% detection rate. Then, to further evaluate the model, we passed the test dataset comprising both attack and normal data instances to other existing models. The results along with the comparison with other recent state-of-the-art models, are presented in Table IV.

After separating a random attack (Infilteration Attack), our attack classifier was trained for 50 epochs and then tested on the test data set. Since the model was not trained for omitted attack class, therefore, it was also removed from the test set at this stage. The overall accuracy attained by the model is illustrated in Table V. In addition to training of BiDLSTM, RF was also trained and evaluated on the same scenario and environment for having a benchmark for comparison.

TABLE IV
A COMPARISON OF PROPOSED AND CURRENT STATE OF THE ART ANOMALY MODELS

| Ref. | Approach | Detection Rate % | Accuracy % | Precision % | False Alarm Rate % |
|------|----------|-----------------|------------|-------------|-------------------|
| [13] | 2L-ZED_IDS AE | 98.7% | 98.3% | 96.9% | 1.08% |
| [39] | ZED-IDS AE | 96.9% | 96.18% | 93.2% | 2.7% |
| [40] | U-Net | 75% | - | 90.6% | - |
| - | Proposed | 100% | 97.8% | 93.9% | 6% |

To evaluate the response of new attacks on our trained models, the omitted attack class was passed into both models. As per RF evaluation, 96.5% of Infilteration attack instances were detected as normal, which was then fed to anomaly detector, which flagged it as an anomaly. Similarly, 94% of the attack was detected as normal by BiDLSTM.

TABLE V
MODEL PERFORMANCE ON 12 CLASSES OF KNOWN ATTACKS

| Approach | Accuracy % | Precision % | Recall % | F1Score % |
|----------|-----------|-------------|----------|-----------|
| BiDLSTM | 92% | 93% | 92% | 92% |
| RF | 96% | 95% | 95% | 95% |

Afterward, these new attack instances were passed to the anomaly detector, which flagged these instances as attacks with 100% accuracy. These new attack instances were mixed in a fixed proportion with old attack instances and normal data by the incremental learning module and then passed to the attack classifier (BiDLSTM) for incremental learning. While for the RF model, these new attack instances were mixed with the old training data, and the model was trained offline from scratch. The difference in recall % of each attack class before and after incremental learning is shown in Table VI

TABLE VI
THE DIFFERENCE IN RECALL% BEFORE AND AFTER INCREMENTAL LEARNING

| Attack Type | Difference in Recall% |
|-------------|----------------------|
| Benign | 1.8% |
| DDOS attack-HOIC | 0% |
| DoS attacks-Hulk | 0.1% |
| Bot | 0.6% |
| DDoS attacks-LOIC-HTTP | 0% |
| SSH-Bruteforce | 0.1% |
| DDOS attack-LOIC-UDP | 0% |
| Brute Force -Web | 0% |
| Brute Force -XSS | 0.9% |
| SQL Injection | 0% |
| DoS attacks-SlowHTTPTest | 0.2% |
| FTP-BruteForce | 0.1% |

The main reason for training another model like RF side by side was to evaluate the incremental learning approach with the offline learning approach. The RF model was trained offline completely with the addition of new attack data into the old training set. The result for RF and the result of the updated model after incremental learning on the same test data set is shown in table VII

TABLE VII
COMPARISON OF INCREMENTAL LEARNING MODEL AND OFFLINE LEARNING MODEL RESULTS AFTER ADDITIONAL ATTACK CLASS

| Approach | Accuracy % | Precision % | Recall % | F1Score % |
|----------|-----------|-------------|----------|-----------|
| BiDLSTM | 91% | 91% | 92% | 92% |
| RF | 95% | 94% | 94% | 95% |

It can be seen from the experiment that the incremental learning model did not lose its performance on old classes. Moreover, it achieved an accuracy of 80% for classifying the new learned attack class. Therefore, adopting incremental learning for this architecture is observed to be the most effective solution. Moreover, to validate the basis of the incremental learning module, some normal data instances were treated as new attack classes and were forwarded to the model for incremental learning. The updated model had an 8% difference in recall of the Benign class, therefore rejecting the updated weights. This extra experiment also validates the logic applied for rejection and acceptance of model weights which was proposed in our architecture.

For achieving better results in terms of classification, different methods proposed in the literature having higher accuracy can be adopted. Similarly, methods proposed for incremental learning showcasing better results can also be implemented in the future to achieve better results than this. Recently, most incremental learning work has been done in the domain of computer vision and image classification. Therefore, models like [17],[25],[41], [42] can be implemented to achieve the desired goal of cybersecurity in UAVs.

## V. CONCLUSION

This research proposes an autonomous self-incremental learning approach for UAVs in the detection of cyber-attacks. Intrusion detectors based on machine learning approaches are widely utilized for detecting and classifying attacks. The two methods utilized in IDS are signature-based and anomaly-based detection. However, both method has their shortcoming. The signature-based detection method cannot distinguish new and unknown attacks, whereas the anomaly detector can although detect new/unknown attacks but is not capable of identifying the type of attack. This gap leads to the need for having an approach capable of identifying both new and old attacks. Keeping this need in mind, many systems have been proposed by combining both of the methods, fulfilling the requirement of detection of unknown attacks. However, if that same attack occurs again, it will be classified as a new attack. This remains a problem to mitigate. To overcome it, we have proposed an architecture capable of identifying new and old attack classes. Our approach uses an incremental learning method for updating signature-based classifiers. Moreover, the proposed architecture is completely independent of humans; that is, it does not require any human interference and can learn continuously on its own.

In this work, BiDLSTM was utilized as a signature-based classifier and LOF was adopted for the anomaly detector. Our LOF model achieves a 100% detection rate for attack instances. Moreover, we have carried out a couple of experiments to cross-validate our proposed architecture and its applicability. In addition to that, a brief overview regarding different incremental learning approaches has also been discussed in this work.

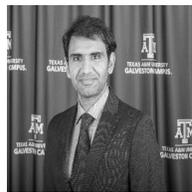

**Irfan Khan** (S'14, M' 18, SM' 20) is an Assistant Professor at the Department of Marine Engineering Technology with a joint appointment with the Electrical and Computer Engineering at Texas A&M College Station. He is the director of the Clean And Resilient Energy Systems (CARES) Lab that focuses on the reliability, sustainability, and security of the electric energy supply on marine vessels. He has been fortunate to receive several grants from multiple funding agencies to work on marine electric distribution systems, electric vehicle fast charging, and electric microgrids. Dr. Khan is an affiliate faculty member with the TAMU Energy Institute and the TEES Smart Grid Center. Before joining TAMU in 2018, Dr. Khan received a Ph.D. in Electrical and Computer Engineering from Carnegie Mellon University USA. He has published more than 90 refereed reputed journal and peer-reviewed conference papers in the smart energy systems-related areas.

Dr. Khan is a registered Professional Engineer (P.E.) with the State of Texas, USA. He is the Vice-Chair for the IEEE Galveston Bay Section (GBS) of Region 5. He has organized several special sessions at various international conferences. Further, Dr. Khan is a regular reviewer of more than 30 reputed journals and conferences, wherein the year 2020, he reviewed more than 230 articles. He is also helping with editorial responsibilities at various journals, e.g., IEEE Transactions on Industry Applications, IEEE Access, Electronics MDPI, Frontiers in Energy Research, etc.

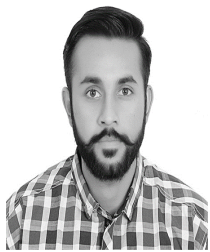

**Yasir Ali Farrukh**, received his B.Eng. in Electrical Engineering from NED University of Engineering and Technology, Karachi, Pakistan, in 2018. Currently, He is pursuing his Ph.D. in Electrical Engineering and working as a research assistant at the Clean and Resilient Energy Systems (CARES) Lab of Texas A&M University, College Station, US.

His current research interest includes Cyber Security, Smart Grid, and AI application technologies.